\documentclass[12pt]{article}
\usepackage{epsfig}

\parskip=\medskipamount 
plus.3em minus.5em \abovedisplayshortskip=.5em plus.2em minus.4em
\renewcommand{\thesection}{\arabic{section}}

\catcode`@=11

\@addtoreset{equation}{section} \@addtoreset{equation}{subsection}
\def\theequation{\ifnum\value{section}=0 \arabic{equation}\ignorespaces
\else \ifnum\value{section}=-1 A.\arabic{equation}\ignorespaces
\else \ifnum\value{subsection}=0
\thesection.\arabic{equation}\ignorespaces \else
\thesection.\arabic{subsection}.\arabic{equation}\ignorespaces
                             \fi
                        \fi
                   \fi}

{\catcode`\'=\active \def'{{}^\bgroup\prim@s}}

\catcode`@=12

\newcommand{\bq}{\begin{equation}}
\newcommand{\be}{\begin{equation}}
\newcommand{\fq}{\end{equation}}
\newcommand{\ee}{\end{equation}}
\newcommand{\bqr}{\begin{eqnarray}}
\newcommand{\beqs}{\begin{eqnarray}}
\newcommand{\fqr}{\end{eqnarray}}
\newcommand{\eeqs}{\end{eqnarray}}

\newcommand{\rf}[1]{(\ref{#1})}

\def\bop#1{\setbox0=\hbox{$#1M$}\mkern1.5mu
    \vbox{\hrule height0pt depth.04\ht0
    \hbox{\vrule width.04\ht0 height.9\ht0 \kern.9\ht0
    \vrule width.04\ht0}\hrule height.04\ht0}\mkern1.5mu}

\def\Prod{\prod}

\begin{document}
\thispagestyle{empty}

\vskip .6in
\begin{center}

{\bf  Realistic Cosmological Constant}

\vskip .6in

{\bf Gordon Chalmers}
\\[5mm]

{e-mail: gordon@quartz.shango.com}

\vskip .5in minus .2in

{\bf Abstract}

\end{center}

Scenarios of supersymmetry breaking at various scales from TeV to GUT to 
the string are generated.  A previous analysis generated the value of 
the experimentally measured cosmological constant from supersymmetry 
breaking at the TeV scale.  Via a reorganization of the perturbative 
series, values of the cosmological constant are generically reconcilable 
with supersymmetry breaking scenarios having scales from the TeV on up 
to the string.  The scenario with only a single supersymmetry 
breaking scale occurs at the GUT scale, generically.

\vfill\break

\noindent{\it Introduction}

The cosmological constant has been assumed to be zero for the prior three 
decades, due possibly to the origins of the work in supersymmetry breaking.  
However, recent astrophysical data indicate that it has a small value, and 
subsequently there have been many attempts to model its value in string 
theory and in quantum field theory.  In fact the value fits the indicative 
formula \cite{Chalmers1}, 

\bqr  
\int {\Lambda^8\over m^4_{\rm pl}} \ , 
\label{cosmo}
\fqr 
with $\Lambda$ approximately the TeV scale (e.g. $1-2$ TeV).  This formula, 
and its origin, is not emphasized much in the literature.  There is also 
an approximate hierarchy to the mass patterns \cite{Chalmers2} of the 
fundamental fermions following from 

\bqr 
\int (\psi^\alpha\psi_\alpha +\psi^{\dot\alpha} \psi_{\dot\alpha}) 
  ~({\Lambda\over m_{\rm pl}})^{n/16} \Lambda \ , 
\label{fermionmasses}
\fqr 
also with $\Lambda$ of the order of a TeV (and $m_{\rm pl}=1.2\times 
10^{28}$), 
and $n$ an integer (the next order correction is presented in 
\cite{Chalmers2}).  The origin of these formulae 
could be explained with supersymmetry breaking at a TeV, however, different  
energies should be more natural given the different scales reflected in 
geometry.   

The nature of the formulae to the patterns of the fundamental masses 
require a deeper explanation \cite{Chalmers2}, possibly in supersymmetry 
breaking \cite{Chalmers3},\cite{Chalmers4}.  In this regard, 
the supersymmetry breaking with differing higher energy scales  
is relevant to the parameters listed in \rf{cosmo} and 
\rf{fermionmasses}, and also to realistic model building.  In 
addition the perturbation theory should be included.  
 
The scenario adopted in \cite{Chalmers1} uses a supersymmetry breaking 
involving eight different scales, $\Lambda_1$ to $\Lambda_8$.  Due to the 
fact that in theory upon three or more scales set to zero the cosmological 
constant equals zero (D-terms in $N=2$ models allow for a classical 
cosmological constant), the functional form of the cosmological 
constant has to be 

\bqr  
{\Lambda^8\over m^4_{\rm pl}} 
 \rightarrow {{\Prod \Lambda_i}\over m_{\rm pl}^4} \ ,
\fqr 
and in the work of \cite{Chalmers1} the scales are taken to be of 
the TeV scale.  A generalization of this work to encompass the string 
scale, as well as energies between the TeV and the string scale, is 
relevant to both particle physics phenomenology and cosmology.

\vskip .2in 
\noindent{\it 'Natural' Value}

The perturbative series expansion contributing to the cosmological 
density is 

\bqr 
V(\Lambda)=\sum b_n \Lambda^4 ({\Lambda^2\over m_{pl}^2})^n \ , 
\label{cosmoexpansion}
\fqr 
with the scale $\Lambda$ in accordance with a supersymmetry breaking 
scale.  For the moment, all supersymmetry breaking scales are taken 
to be identical and equal to $\Lambda$.  The coefficients $b_n$ are 
determined by the loop expansion.  This expansion generically produces 
a $V(\Lambda)$ which is $10^{56}$ greater than the observed value, if 
the TeV scale is chosen for $\Lambda$.  

A re-ordering of the perturbative series can be easily implemented.  
For example, the first two terms in the series are grouped in the 
manner, 

\bqr 
b_0 \Lambda^4 + b_1 {\Lambda^6/ m_{pl}^2} + \ldots = 
 c_0 \Lambda^2 m_{pl}^2 ~(e^{-c_1 \Lambda^2/m_{\rm pl}^2}-1) + 
\ldots  \ ,
\label{reordering}
\fqr 
so that the expansion of the exponential generates \rf{cosmoexpansion}.  
The remainder terms are altered in accordance to agree with 
\rf{cosmoexpansion}; they may also be written in terms of exponentials.  
This rewriting of the series can be done to any order of accuracy.  

Physically, there is no reason a priori to suspect that either the 
form in \rf{cosmoexpansion} or \rf{reordering} is more physical, aside 
from the physical meaning of the exponential which could be interpreted 
as an instanton-like effect.  

The cosmological data is next used to determined the value of $\Lambda$. 
The powers of the scale breaking are determined by $\Lambda=10^x$, and 
$e^{10}=10^{4.34}$, as 

\bqr 
8.68(x-28)+2x+56=-8 
\qquad 
10.7x=179 \ .  
\fqr 
In the units of this paper, the 'accepted' value of the observed 
cosmological constant is $10^{-8}$.  
This calculation results in the scale of supersymmetry breaking being 
\bqr 
x=15.5  \qquad \Lambda=10^{16.7} \ .  
\fqr 
The factors of $100$ to $1000$ in the coefficients $c_0$ and $c_1$ 
change little the $x$ value ($16-17$).  The scale of $\Lambda$ is the 
Grand Unified Theory (GUT) scale.  This is interesting, as a supersymmetry 
breaking scale at the GUT scale will generate the currently 'accepted' 
non-vanishing value of the cosmological constant. 

The next term in the series has the value, 
\bqr 
{\Lambda^8\over m_{\rm pl}^4} \sim 10^{136-112}=10^{18}
\fqr 
and is $28$ orders of magnitude too large.  Another exponential rewriting 
can be performed to lower its value in accordance with the previous two 
terms.  Then there is a relative suppression of $\Lambda^2/m_{pl}^2\sim 
10^{-22}$.  Otherwise, fine tuning of the coefficient $b_2$ is required 
to an accuracy of $28$ digits.  The $\Lambda^{10}/m_{\rm pl}^4$ term is 
of order $10^{3}$ and requires exponention; this term can be packaged 
with the previous one.  

In comparison, the following terms are shown to be irrelevant (to this 
accuracy) in either case, with reordering or without reordering of the 
series \rf{cosmoexpansion}.  The term $\Lambda^{12}/m_{pl}^8=10^{-20}$ 
for $x=17$, and subsequent ones are of smaller value.

In effect, the rewriting of the perturbative series in the form 
\rf{reordering} naturally predicts a scale of $10^{16-17}$, the GUT 
scale, for the supersymmetry breaking.  This prediction is very stable 
under perturbations of the coefficients $b_n$.

\vskip .2in 
\noindent{\it Two Scales}

Various degrees of supersymmetry breaking are expected to happen at 
different scales.  The multiple scale scenario can generate breakings 
at the string scale combined with breaking at the TeV scale.  These 
cases are considered next.  The predictions are interesting in that 
it straightforward to obtain scales in the GeV range on up to the string 
energy.  

First consider the scenario of two scales $\Lambda_1$ and $\Lambda_2$.
In the absence of a potential allowable D-term, which contributes at 
the classical level a term $\Lambda^6/m_{\rm pl}^2$, the cosmological 
constant should vanish upon taking the scales to zero.  This means 
that in the termwise, taking $\Lambda_1\rightarrow 0$ or 
$\Lambda_2\rightarrow 0$, the function to $V(\Lambda_1,\Lambda_2)$ must 
go to zero.   For this, the function is modeled by, 

\bqr 
 V= b_0 (\Lambda_1\Lambda_2)^2 + b_1 (\Lambda_1\Lambda_2)^3/m_{\rm pl}^2 + 
 \ldots \ . 
\label{twoscales} 
\fqr 
The expansion in \rf{twoscales} depends on there being two scales, and 
it is indicative of the two scale supersymmetry breaking scenario.  

The reordering of the series in \rf{twoscales} can be done in a couple 
of ways.  The first one considered, and with $\Lambda_2>\Lambda_1$, 
follows by rewriting the first two terms as, 

\bqr 
c_1 \Lambda_1\Lambda_2 m_{pl}^2 (e^{-c_2 \Lambda_1^2/m_{\rm pl}^2}-1) \ .  
\fqr 
Consider the case when $\Lambda_2=\Lambda_1^2$.  The effects of 
the exponent are, with $\Lambda_1=10^x$, 

\bqr 
8.68(x-28)+3x+56=-8  \qquad 11.68x=179
\fqr 
and generates, 

\bqr 
x=15.3  \ .  
\fqr 
This is unphysical as the lower scale is an the order of a $1000$ TeV, 
but the higher scale is at $10^{30}$ eV, which is higher than the string 
scale.  

The next possibility is use the higher scale in the exponent.  This 
results in the numbers, 

\bqr 
8.68(2x-28)+3x+56=-8  \qquad x=9.1 \ .  
\fqr 
In this case, supersymmetry breaking occurs at a $1$ GeV and at the 
approximate GUT scale ($10^{18}$).  

The final two scale scenario considered uses the exponential, 

\bqr 
e^{-\lambda_1\lambda_2/m_{\rm pl}^2} \ , 
\fqr 
and results in the numbers, 

\bqr 
4.34 (x-28)+4.34 (2x-28)+3x+56=-8  
\fqr 
or, 
\bqr 
16x=179  \qquad x=11.2  \ .  
\fqr 
The supersymmetry breakings are near a TeV and at also at the string 
scale.  

The two scale scenario shows that it is possible to obtain, with the 
rewriting of the series, supersymmetry breaking at very different 
scales.  Namely, it is possible to obtain the two breaking such that 
one is near the TeV scale, or in the $100$ GeV scale, and the other is 
near the string scale, or in the GUT scale regime.  This happens 
with a realistic prediction of the cosmological constant.  

It should be noted that the higher order terms, such as 
$\Lambda^8/m_{\rm pl}^4$, should also be exponentiated.  The higher 
order terms in the case of $\Lambda=10^{11}$ are, 

\bqr 
\lambda_1^4\lambda_2^4/m_{pl}^4  \qquad  10^{44+88-112}=10^{20}  
\fqr 
\bqr \hskip .08in 
\lambda_1^5\lambda_2^5/m_{pl}^6  \qquad  10^{55+110-168}=10^{-3}  
\fqr
\bqr \hskip .12in
\lambda_1^6\lambda_2^6/m_{pl}^8  \qquad  10^{66+132-224}=10^{-26} \ . 
\fqr 
The first and second terms requires the exponentation, as the first is $28$ orders 
of magnitude too big, and the remaining terms are in agreeement with 
the $10^{-8}$, the 'accepted' observed value of the cosmological 
constant in these units. 

\vskip .2in 
\noindent{\it Three Scales}  

The supersymmetry breaking with three scales also has some ambiguity 
in the rewriting of the perturbative series, and allows for much space 
between the energy scales.  Consider the series for 
$V(\Lambda_i)$ of the form, 

\bqr 
V(\Lambda_i)=b_0 \sum_{i=1}^4 \Lambda_1\Lambda_2\Lambda_3\Lambda_i 
+ b_1 (\Lambda_1\Lambda_2\Lambda_3)^2/m_{\rm pl}^2 + \ldots \ ,  
\fqr 
with $\Lambda_3>\Lambda_2>\Lambda_1$.  

Upon rewriting the first two terms as 

\bqr 
\sum_{i=1}^4 c_0 \Lambda_3\Lambda_i m_{\rm pl}^2 
  ~(e^{-c_1 \Lambda_1\Lambda_2/m_{\rm pl}^2}-1) \ , 
\label{threeone}
\fqr 
the counting of the exponents with $i=3$ (the largest contribution) 
generates with $\Lambda_i=10^{x_i}$, 

\bqr 
2x_3+56+4.34(x_1-28)+4.34(x_2-28)=-8  \ .  
\fqr 
This equation is not ambiguous, but gives a scale relation between 
the $\Lambda_i$.  Note that the form in \rf{threeone} does not have 
the quartet term matching, but rather is a bound because the largest 
symmetry breaking scale $\Lambda_3$ is used; a factor of three in 
the coefficient $c_0$ absorbs this, but other forms may also be used.

Consider the example of $x_1=12$ ($\Lambda=10^{12}$) and $x_2=x_3=x$.   
The relevant numbers are, 

\bqr 
6.34x=127  \qquad x=20 \ .  
\fqr 
In this case $\Lambda_1=10^{12}$ which is the TeV scale, and two more 
closer to the string scale at $\Lambda_2=\Lambda_3=10^{20}$.  The latter 
should be spaced a bit as three scales are being considered.   

The second scenario is 
\bqr 
\sum_{i=1}^4 c_0 \Lambda_1\Lambda_i m_{\rm pl}^2 
  ~(e^{-c_1 \Lambda_2\Lambda_3/m_{\rm pl}^2}-1) \ , 
\fqr 
the counting of the exponents with $i=3$ (the largest contribution) 
generates, 

\bqr 
x_1+x_3+56+4.34(x_2-28)+4.34(x_3-28)=-8  \ .     
\fqr 
Again, the quartet term can be bounded by a factor of three due to 
the choice of exponentiation.  
With the same breaking, $x_1=12$ and $x_2=x_3=x$, the numbers are, 

\bqr 
9.68x=167 \qquad x=17 \ .  
\fqr 
The three scales are then at the TeV scale together with another two 
at the GUT scale.  There is considerable flexibility though in the examples, 
when two of the scales are not chosen together.   

To demonstrate the flexibility consider the final example, with $x_1=12$ 
and $x_3={3\over 2} x_2$.  Take the first rewriting in the previous, and the 
numbers are 

\bqr 
12.4 x_2 =167 \qquad x_2=14 \ .  
\fqr 
In this case, $\Lambda_1=10^{12}$, $\Lambda_2=10^{14}$, and 
$\Lambda_3=10^{21}$.  The scales are at a TeV, the GUT energy, and 
also at the string scale.  Again, there is much freedom in arranging 
the scales, and there are potentially other interesting cases.  

In the previous examples, the higher order terms must also be arranged 
in exponential form in order to not spoil the value of the cosmological 
constant.  It is clear, however, that scenarious with various scales 
of disparate supersymmetry breaking and that are realistic, may be 
obtained.  This is without fine-tuning, and without any real model 
specificity, unless actual values of the supersymmetry breaking 
are used.

\vskip .2in 
\noindent{\it Conclusion}

The cosmological constant problem is typically very difficult to 
solve without some fine-tuning.  However, it is possible to use a 
simple 'rearrangement' of the perturbative series to obtain realistic 
values, at least in accord with the current supernovae data.  

The simplest scenario, requiring only one supersymmetry breaking scale, 
indicates that with the current 'accepted' value of the cosmological 
constant, that the supersymmetry breaking is at the scale of $10^{16-17}$ 
eV.  This is interesting for a variety of reasons, not only because the 
prediction is generic, but also due to the one-loop coupling unification 
of the standard model at this scale.

More interesting scenarios of supersymmetry breaking can be achieved if the  
models possess more than one scale.  Three or four scales can be used 
in a direct fashion to obtain supersymmetry breaking at the TeV scale, 
the GUT scale, and the string scale, in one model.  The typical fine-tuning 
problem of supersymmetry breaking is compounded many fold with supersymmetry 
breaking at the string scale.  This work shows that no real 
fine-tuning is required even with scales of supersymmetry breaking located 
near the string as well as in current phenomenological experiments. 

The dynamical origin of the possible running of the scales would be 
interesting to further understand due to the perturbative series 
used in this work.

\vfill\break

\end{document}